\documentclass[dvipsnames,twocolumn]{aastex631}
\usepackage{graphicx}
\usepackage{amsmath}
\usepackage{txfonts}
\usepackage{color}

\newcommand{\mdisk}[0]{\ensuremath{M_{\rm disk}}}
\newcommand{\mgas}[0]{\ensuremath{M_{\rm gas}}}

\newcommand{\msun}[0]{\ensuremath{\mathrm{M}_{\odot}}}

\newcommand{\nhp}[0]{\ensuremath{\mathrm{N}_2\mathrm{H}^+}}
\newcommand{\co}[0]{\ensuremath{^{12}\mathrm{CO}}}
\newcommand{\xco}[0]{\ensuremath{^{13}\mathrm{CO}}}
\newcommand{\cyo}[0]{\ensuremath{\mathrm{C}^{18}\mathrm{O}}}
\newcommand{\zet}[0]{\ensuremath{\zeta_{\rm CR}}}
\newcommand{\htwo}[0]{\ensuremath{\mathrm{H}_{2}}}
\newcommand{\abu}[0]{\ensuremath{x_{\rm CO}}}

\begin{document}
\title{A novel way of measuring the gas disk mass of protoplanetary disks using \nhp\ and \cyo}

\correspondingauthor{Leon Trapman}
\email{ltrapman@wisc.edu}
\author[0000-0002-8623-9703]{Leon Trapman}
\affiliation{Department of Astronomy, University of Wisconsin-Madison, 
475 N Charter St, Madison, WI 53706, USA}

\author[0000-0002-0661-7517]{Ke Zhang}
\affiliation{Department of Astronomy, University of Wisconsin-Madison, 
475 N Charter St, Madison, WI 53706, USA}

\author[0000-0002-2555-9869]{Merel L. R. van 't Hoff}
\affiliation{Department of Astronomy, University of Michigan,
323 West Hall, 1085 S. University Avenue,
Ann Arbor, MI 48109, USA}

\author[0000-0001-5217-537X]{Michiel R. Hogerheijde}
\affiliation{Leiden Observatory, Leiden University, P.O. box 9513, 2300 RA Leiden, The Netherlands}
\affiliation{Anton Pannekoek Institute for Astronomy, University of Amsterdam, Science Park 904, 1098 XH Amsterdam, The Netherlands}

\author[0000-0003-4179-6394]{Edwin A. Bergin}
\affiliation{Department of Astronomy, University of Michigan,
323 West Hall, 1085 S. University Avenue,
Ann Arbor, MI 48109, USA}

\begin{abstract}
Measuring the gas mass of protoplanetary disks, the reservoir available for giant planet formation, has proven to be difficult. 
We currently lack a far-infrared observatory capable of observing HD, and 
the most common gas mass tracer, CO, suffers from a poorly constrained CO-to-H$_2$ ratio. 
Expanding on previous work, we investigate if \nhp, a chemical tracer of CO poor gas, can be used to observationally measure the CO-to-H$_2$ ratio and correct CO-based gas masses.
Using disk structures obtained from the literature, we set up thermochemical models for three disks, TW Hya, DM Tau and GM Aur, to examine how well the CO-to-H$_2$ ratio and gas mass can be measured from \nhp\ and \cyo\ line fluxes.
Furthermore, we compare these gas masses to independently gas masses measured from archival HD observations.
The \nhp\,(3-2)/\cyo\,(2-1) line ratio scales with the disk CO-to-H$_2$ ratio.
Using these two lines, we measure $4.6\times10^{-3}\ \msun \leq \mdisk \leq 1.1\times10^{-1}\ \msun$ for TW Hya, $1.5\times10^{-2}\ \msun\leq \mdisk \leq 9.6\times10^{-2}\ \msun$ for GM Aur and $3.1\times10^{-2}\ \msun \leq \mdisk \leq 9.6\times10^{-2} \msun$ for DM Tau.
These gas masses agree with values obtained from HD within their respective uncertainties. 
The uncertainty on the \nhp\,+\,\cyo\ gas mass can be reduced by observationally constraining the cosmic ray ionization rate in disks.
These results demonstrate the potential of using the combination of \nhp\ and \cyo\ to measure gas masses of protoplanetary disks.
\end{abstract}

%
\section{Introduction}
\label{sec: introduction}

The gas mass of protoplanetary disks is a crucial ingredient for planet formation theories (e.g. \citealt{Mordasini2018}). On a macro scale, the gas mass represents the total mass budget available for forming gas giants. Combined with the stellar accretion rate and mass loss rate it determines the lifetime of the disk and therefore sets the timescale for giant planet formation.
On a micro scale, the gas density and gas-to-dust ratio regulate the dynamics of dust grains and larger bodies in the disk. The rate at which the dust grows, settles towards the midplane and drift inward toward the central star all depend on how much gas is present in the disk (see, e.g., \citealt{Birnstiel2012}). 

Measuring the gas mass of disks from observations has proven difficult. The gas is predominantly made up of molecular hydrogen (\htwo), a light, symmetric molecule without a permanent dipole moment that does not significantly emit at temperatures typically found in protoplanetary disks. 
Hydrogen deuteride (HD) has been shown to be a promising indirect tracer of the gas mass. HD is chemically similar to \htwo\ and thus closely follows the distribution of \htwo\ (see \citealt{Trapman2017}). Contrary to \htwo, HD has a small dipole moment and can emit from a significant part of the disk. 
Using \textit{Hershel}, the HD $J=1-0$ line has been detected in three protoplanetary disks: TW Hya, GM Aur and DM Tau \citep{Bergin2013, McClure2016}, allowing us to accurately measure their gas masses. HD 1-0 upper limits have also been used to put constraints on the gas masses of Herbig disks \citep{Kama2020}. With the end of the \textit{Herschel} mission, we currently lack a far-infrared observatory capable of observing HD in disks.

Instead, the most often used gas mass tracer is carbon monoxide (CO), the second most abundant molecule in the gas of protoplanetary disks. CO emission is bright at millimeter wavelengths and has been detected in almost all protoplanetary disks.  
The emission of its main isotopologue, \co, is optically thick in disks, but its less abundant isotopologues, e.g. \xco\ and \cyo, are most often optically thin.
Emission from rarer isotopologues like $^{13}$C$^{18}$O or $^{13}$C$^{17}$O is even more optically thin, but these weak lines are less suitable for large gas mass surveys (see e.g. \citealt{Zhang2017,Booth2019}).
The main uncertainty when using CO as gas mass tracer is the CO-to-\htwo\ abundance ratio (\abu). CO is a chemically stable molecule that in the warm gas of protoplanetary disks is expected to have a relatively constant abundance of $\abu\approx10^{-4}$. Two main processes reduce \abu\ in the disk.
In the irradiated surface layer of the disk far-ultraviolet photons will photo-dissociated CO until it can build up a sufficient column to shield itself against this process. 
Close to the cold midplane of the disk CO freezes out onto the surface of dust grains. Both of these processes have been extensively studied and are well understood (see, e.g. \citealt{vanDishoeckBlack1988,Zadelhoff2001,Visser2009,Miotello2014}). 
By incorporating them in physical-chemical disk models these processes can be accounted for, allowing us to measure disk masses from \xco\ and \cyo\ line fluxes (e.g. \citealt{WilliamsBest2014,Miotello2016}).

However, there is increasing observational evidence that we are overlooking one or more processes that reduce \abu\ in protoplanetary disks. When compared to gas masses derived independently from HD, the CO-based gas masses are found to be $5\,-\,100\,\times$ lower, even after including the effects of photo-dissociation and freeze-out (see, e.g. \citealt{Favre2013,Schwarz2016,Kama2016,McClure2016,Trapman2017,Calahan2021}). This could imply that the low CO-based gas masses found in disk surveys carried out with the Atacama Large Millimiter/submillimeter Array (ALMA) are severely underestimating the gas masses of protoplanetary disks (see, e.g. \citealt{ansdell2016,miotello2017,Long2017}). Several processes have been suggested to cause this low \abu, such as the chemical conversion of CO into more complex species (e.g. \citealt{Aikawa1997,FuruyaAikawa2014,Yu2016,Yu2017,Bosman2018b,Schwarz2018}). Alternatively, grain growth can lock up CO into larger dust bodies that settle towards the midplane (e.g. \citealt{Bergin2010, Bergin2016,Kama2016,Krijt2018,Krijt2020}). 
Irrespective of what causes the low CO abundance, we need to measure \abu\ in order to reliably measure disk gas masses using CO.

In this Letter we examine \nhp, a chemical tracer of CO-poor gas, as a potential calibrator of the CO abundance in protoplanetary disks. 
\nhp\ is formed through proton transfer from H$_3^+$ to N$_2$. If CO is present in the gas phase it competes with N$_2$ for the available H$_3^+$, thus impeding the formation of \nhp. In addition, \nhp\ is rapidly destroyed by gas-phase CO. As a result, \nhp\ is therefore only abundant if there is a lack of gas-phase CO.
These properties have seen \nhp\ be successfully used to locate the CO iceline (e.g. \citealt{Qi2013,Qi2015,Qi2019,Huang2015,vtHoff2017}). 

\begin{figure}
    \centering
    \includegraphics[width=\columnwidth]{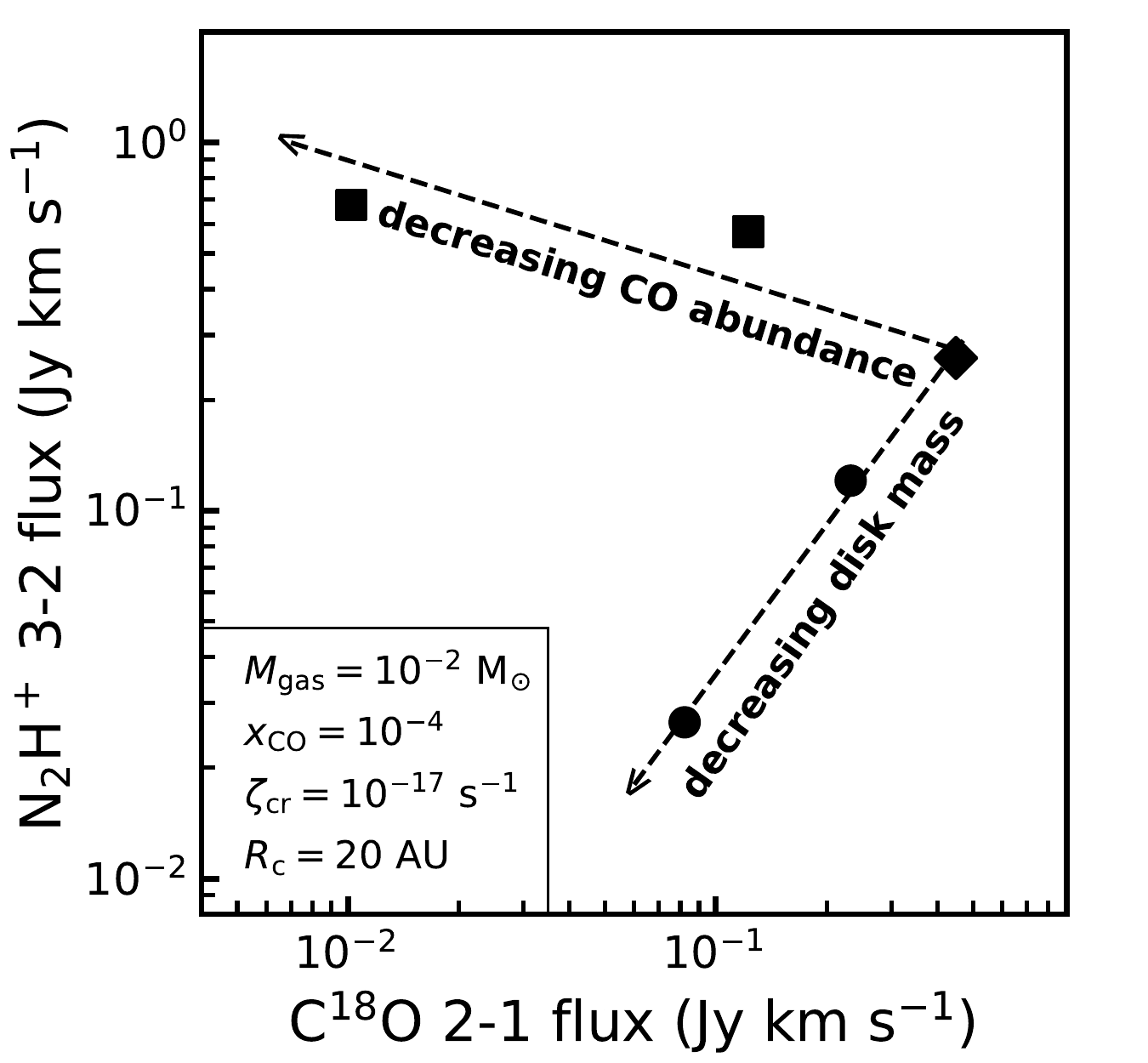}
    \caption{\label{fig: concept Mass vs CO abu} Concept of how \nhp\ and \cyo\ enable us to distinguish between disks with low disk masses and low CO abundance. Starting from the fiducial model (diamond) on the right, points show the change in \nhp\ 3-2 and \cyo\ 2-1 line flux when either the disk mass (circles) or CO abundance (squares) is decreased, with decrements of factors of ten. Relevant disk parameters of the fiducial model are shown in the bottom left. }
\end{figure}

For the disk as a whole, this chemical relation between \nhp\ and CO means higher \nhp\ abundances, and hence higher fluxes, when CO is underabundant
(see Figure \ref{fig: concept Mass vs CO abu}).
The anti-correlation between the \nhp\ flux and the CO abundance was previously studied in \cite{Anderson2019}, who showed that a CO abundance $\abu\leq10^{-6}$ is required to explain the \nhp\ and CO line fluxes of two disks in Upper Sco.
Here we expand upon their work and examine if \nhp\ can be used to observationally measure the global CO abundance in protoplanetary disks and therefore, the gas mass. To test this hypothesis we select the three disks with HD $J=1-0$ detections, TW Hya, DM Tau and GM Aur, for which we have independent measurements of the disk gas mass and global CO abundance. 

The structure of this Letter is as follows:
in Section \ref{sec: model setup} we set up thermochemical disk models based on disk structures obtained from the literature and use the \cyo\ $J=2-1$ integrated flux to select the combinations of CO abundance and disk gas mass that reproduce the \cyo\ observations. 
Combining a simple \nhp\ chemical network with our thermochemical models to calculate \nhp\ abundances and excitation, we show in Section \ref{sec: results} that the \nhp\ $J=3-2$ integrated line flux, combined with the \cyo\ 2-1 flux, constrains the average CO abundance in the disk. For a reasonable range of cosmic ray ionization rates we show that correcting the CO-based gas mass using \abu\ derived from \nhp\ and \cyo\ agrees with the gas disk mass measured from HD.
In Section \ref{sec: discussion} we examine the cosmic ray ionization rates in our three disks and we discuss some of the caveats of this study.
We summarize our findings in Section \ref{sec: conclusions}.

\begin{table*}[th]
\centering
\caption{Source information}
\begin{center}
\begin{tabular}{lcccccccccccc}
\hline
\hline
Source & $M_{\star}$ & $L_{\star}$ &  T$_{\rm eff}$ & $L_{\rm UV}$ & $L_X$ & dist & $M_{\rm gas}$ (HD) & $M_{\rm gas}$ (CO)  & \nhp\ $3-2$ & \cyo\ $2-1$ & HD $1-0$ \\
& (M$_{\odot}$) & (L$_{\odot}$) & (K) & (erg s$^{-1}$) & (erg s$^{-1}$) & (pc) & (M$_{\odot}$) & (M$_{\odot}$) & (Jy km/s) & (Jy km/s) & (W m$^{-2}$)  \\
 & & & & $\times10^{30}$ & $\times10^{30}$ & & $\times10^{-2}$ & $\times10^{-2}$ &  & & $\times10^{-18}$ \\
\hline
TW Hya & 0.8 & 0.28& 4110& $27$ & $1.6$ & 59  & 0.77-2.5&  0.05-0.5   &   $2.0\pm0.21$   & $0.57\pm0.06$ & $6.3\pm0.7$\\
DM Tau & 0.53&  0.24& 3705& $3.0$  & $0.3$  & 145 & 1-4.7   &  0.1-1 &  $1.286\pm0.17$ & $1.11\pm0.11$ & $1.6\pm0.4$ \\
GM Aur & 1.1 & 1.2 & 4350& $28$  & $1.4$ & 159 & 20     & 0.8 &  $1.487\pm0.18$ & $1.024\pm0.097$ & $2.5\pm0.5$\\
\hline
 Refs & \multicolumn{3}{c}{(1,2,3)} & (4,5,6) & (7,6) & (3) & (8,9,10,11) & (12,13,14,15) & (16,17) & (9,18,19) & (20,10) \\
\end{tabular}
\label{tab:source_info}
\end{center}
\begin{minipage}{0.95\textwidth}
\vspace{-0.25cm}
{\em Notes}: for details on the adopted disk structures, see \cite{Kama2016} for TW Hya,           \cite{Zhang2019} for DM Tau and \cite{ZhangMAPS2021} for GM Aur. ALMA line fluxes include a 10\% systematic flux uncertainty.

          \em{References: (1) \cite{Andrews2012}, 
                          (2) \cite{KenyonHartmann1987},
                          (3) \cite{GaiaDR2_2018},
                          (4) \cite{Cleeves2015},
                          (5) \cite{Dionatos2019}
                          (6) \cite{Brickhouse2010},
                          (7) \cite{Henning2010}
                          (8) \cite{Trapman2017},
                          (9) \cite{Calahan2021},
                         (10) \cite{McClure2016},
                         (11) \cite{SchwarzMAPS2021},
                         (12) \cite{Thi2010},
                         (13) \cite{Miotello2016},
                         (14) \cite{Zhang2019}, 
                         (15) \cite{ZhangMAPS2021},
                         (16) Qi et al., in prep.,
                         (17) \cite{Qi2019},
                         (18) \cite{Bergner2019},
                         (19) \cite{ObergMAPS2021},
                         (20) \cite{Bergin2013}
                          }
    \end{minipage}
\end{table*}

\section{Model setup}
\label{sec: model setup}

For running our models we use the thermochemical code \underline{D}ust \underline{a}nd \underline{Li}nes (\texttt{DALI}; \citealt{Bruderer2012,Bruderer2013}). For a physical 2D disk structure and stellar spectrum \texttt{DALI} calculates the thermal and chemical structure of the disk self-consistently. The computation is split into three steps: first, the radiative transfer equation is solved using a 2D Monte Carlo method to calculate the dust temperature structure and the internal radiation field. 
Next, the abundances of molecular and atomic species are calculated at each point in the disk by solving the time-dependent chemistry. We include CO and HD isotope-selective photodissociation and chemistry following \cite{Miotello2014} and \cite{Trapman2017}. 
The excitation levels of all species are computed using a non-LTE calculation. From the excitation levels the gas temperature is calculated by balancing heating and cooling processes. As both the chemistry and excitation depend on the gas temperature, an iterative calculation is used to find a self-consistent solution. 
Finally, the model is ray-traced to produce integrated line fluxes and line profiles. For a more detailed description of the code, see Appendix A of \cite{Bruderer2012}. 

\begin{figure}
    \centering
    \includegraphics[width=\columnwidth]{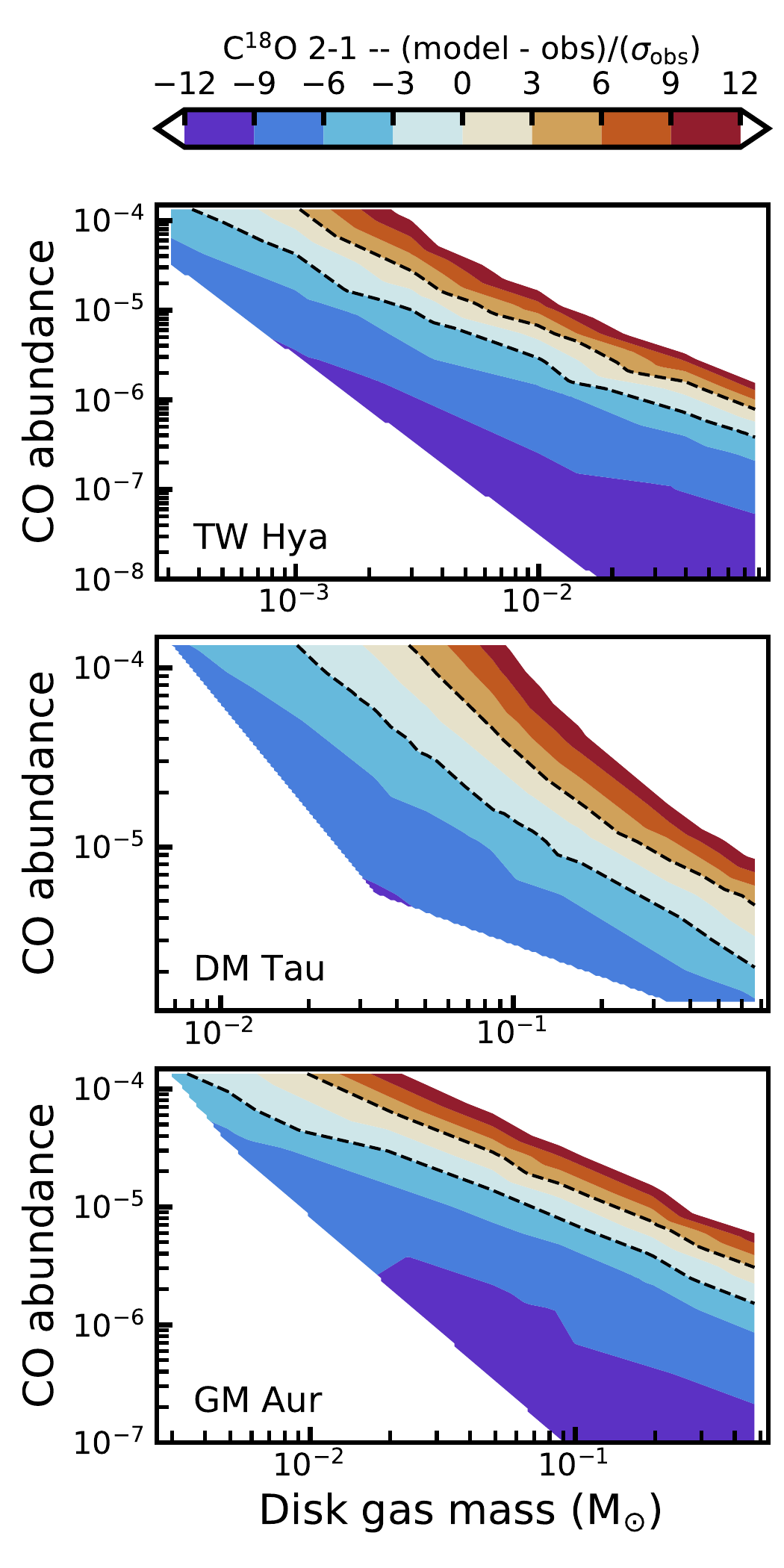}
    \caption{\label{fig: C18O score} Comparison between the observed \cyo\ $J=2-1$ integrated line flux and our models with different combinations of disk gas mass and CO abundance. Colors show how close the model \cyo\ line flux is to the observations, expressed in terms of the uncertainty on the \cyo\ line flux. $(\sigma_{\rm obs})$. Dashed black lines highlight the part of the \mgas-\abu\ parameters space where model \cyo\ fluxes are within $3 \sigma_{\rm obs}$. Darker shades of blue (red) marks parameter space where the model underproduces (overproduces) the observed flux.  }
\end{figure}

For the three disks with HD $J=1-0$ detections, we obtain disk structures from the literature (TW Hya: \citealt{Kama2016}; DM Tau: \citealt{Zhang2019}; GM Aur: \citealt{ZhangMAPS2021}). For each of these models, the disk structure was fitted using a combination of the spectral energy density (SED) and resolved millimeter continuum and line emission. 
With these fiducial models our approach is as follows: we set up and run a set of models with different combinations of $M_{\rm gas}$ and $\abu$. From this set of models we find the subset that reproduces the integrated $\cyo\ J=2-1$ flux. Specifically we select those models that satisfy $F_{\cyo}^{\rm model} - F_{\cyo}^{\rm obs} \leq 3 \sigma_{\cyo}^{\rm obs}$. Our models show that the $\cyo\ J=2-1$ flux is mostly optically thin, only becoming optically thick in the inner disk $(R<0.5-1\times R_c)$.
Note that this inner disk area encompasses $40-60\%$ of the total gas mass in each disk.
These models represent the combinations of gas disk mass and CO abundance that could explain the \cyo\ observations. Models where $\abu< 10^{-4}$ represent scenarios where some amount of the gas-phase CO is either chemically converted or locked up as ice on large dust grains. By lowering the CO abundance in this way we remain agnostic on how the CO is removed from the gas and focus on the end result, an underabundance of gaseous CO in disks.
The model selection is visualized in Figure \ref{fig: C18O score}.

For the models that reproduce the observed \cyo\ flux, we calculate the \nhp\ chemistry using the chemical network presented in \cite{vtHoff2017}. Briefly, this network includes the production of H$_3^+$ through cosmic-ray induced ionization of H$_2$, freeze-out and desorption of CO and N$_2$, formation of \nhp\ and HCO$^+$ through reactions of H$_3^+$ with N$_2$ and CO, respectively, and the destruction of \nhp\ through reacting with CO (see their Figure 2). 
Using this simple network instead of the larger network in DALI has the advantage of being easier to understand what factors could affect the \nhp\ abundance and fluxes (see also Section \ref{sec: constraining the cosmic ray ionization rate}). 

Our approach is the following: we take the gas temperature structure and CO and N$_2$ abundance calculated with DALI and use these to compute the \nhp\ abundance structure using the chemical \nhp\ network. The main free parameter in this network is the cosmic ray ionization rate \zet. We note that \zet\ is the only source of ionization in our network. While cosmic rays are likely the dominant ionization source in the region of the disk where \nhp\ is abundant (e.g. \citealt{Aikawa2015,AikawaMAPS2021}), the values of \zet\ in this work should be read as upper limits, as other ionization processes such as X-rays and the decay of radionuclids could still contribute (see e.g. \citealt{Seifert2021}).  
We compute the \nhp\ abundance structure for $\zet = 10^{-19}, 10^{-18}\ \mathrm{and}\ {10^{-17}\ \mathrm{s}^{-1}}$.
The \nhp\ abundance structure is then re-inserted into the DALI model to calculate the excitation and synthetic line emission.

\section{Results}
\label{sec: results}

\begin{figure}
    \centering
    \begin{minipage}{0.99\columnwidth}
    \includegraphics[width=\columnwidth]{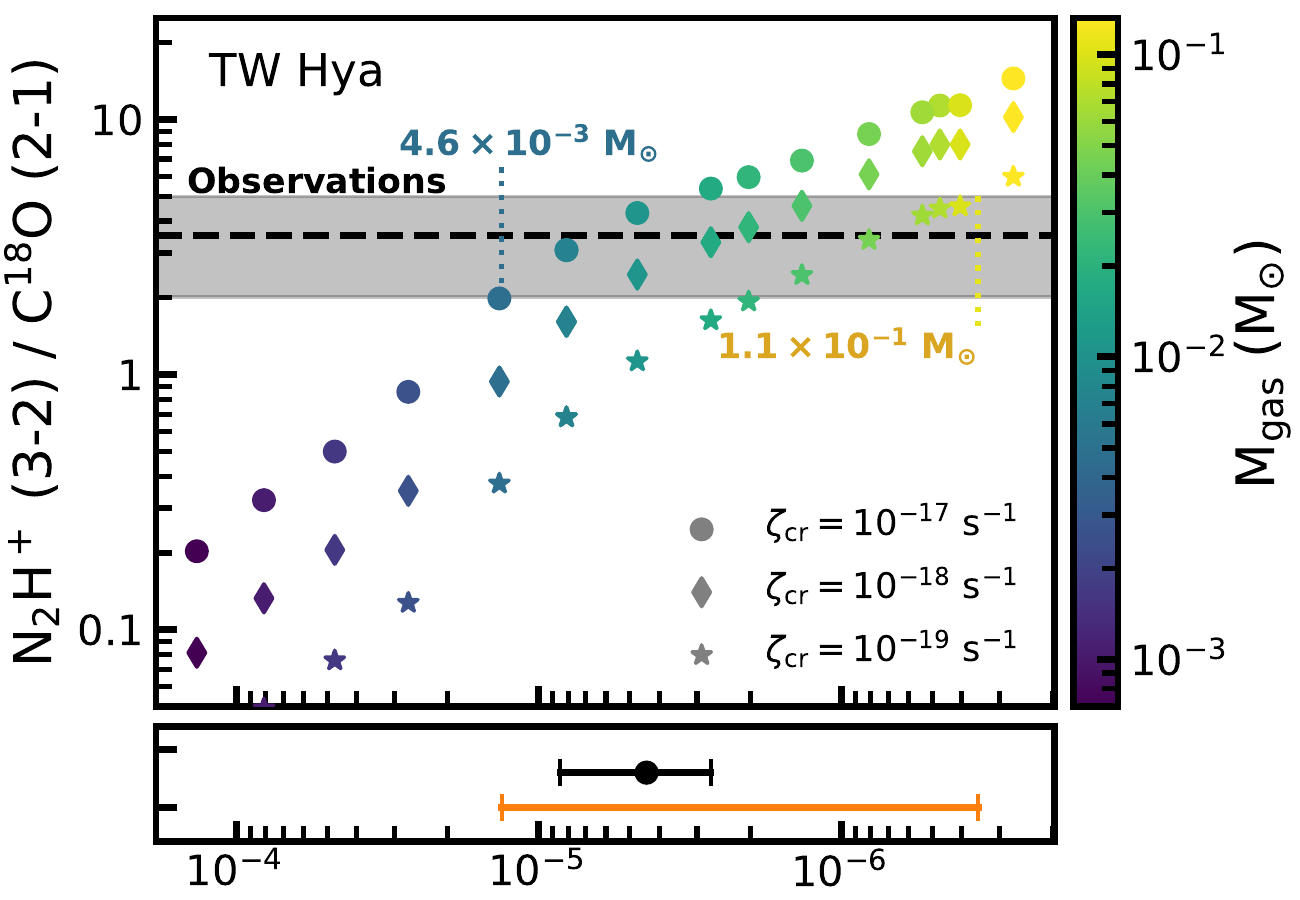}
    \end{minipage}
    \begin{minipage}{0.99\columnwidth}
    \includegraphics[width=\columnwidth]{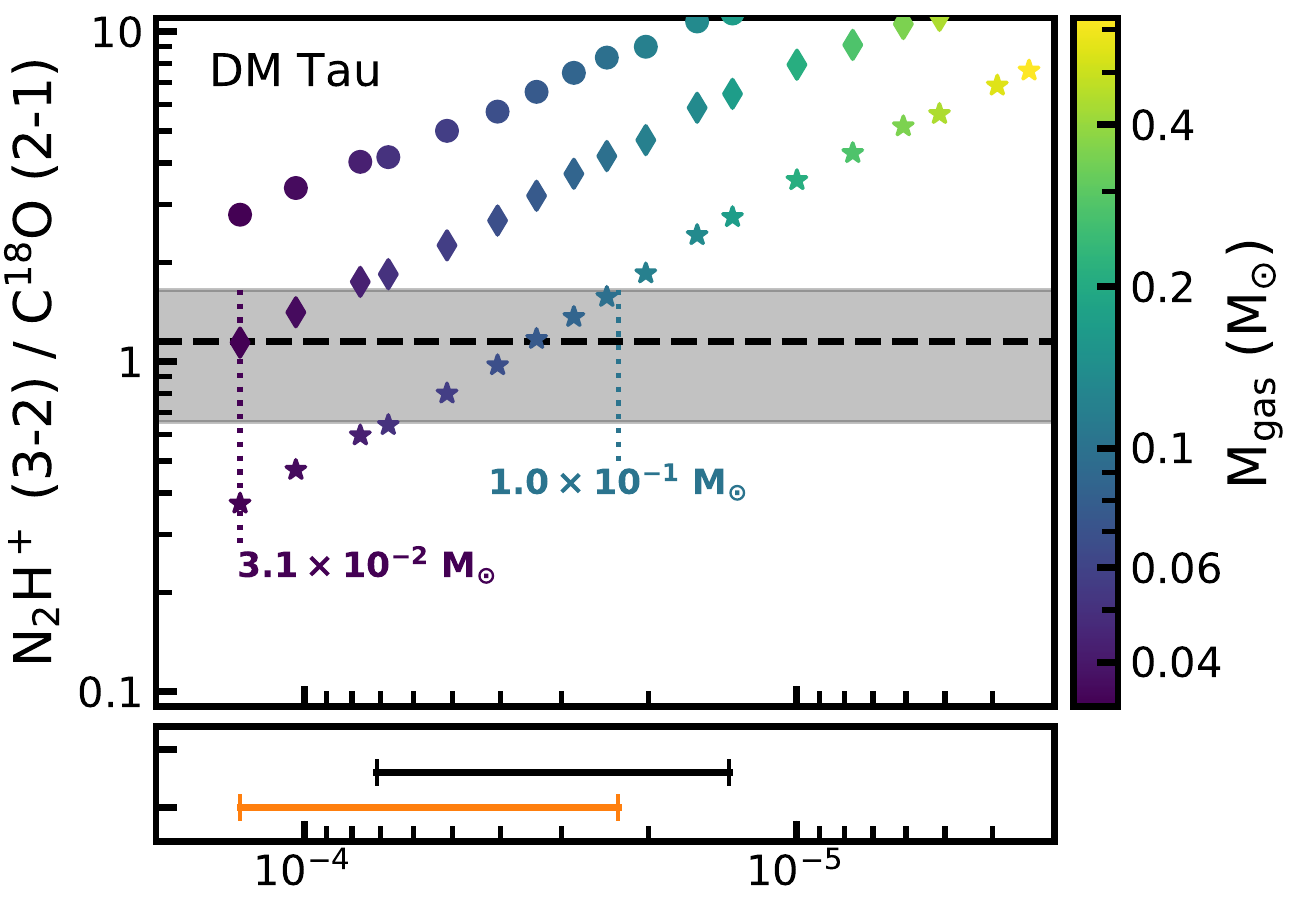}
    \end{minipage}
    \begin{minipage}{0.99\columnwidth}
    \includegraphics[width=\columnwidth]{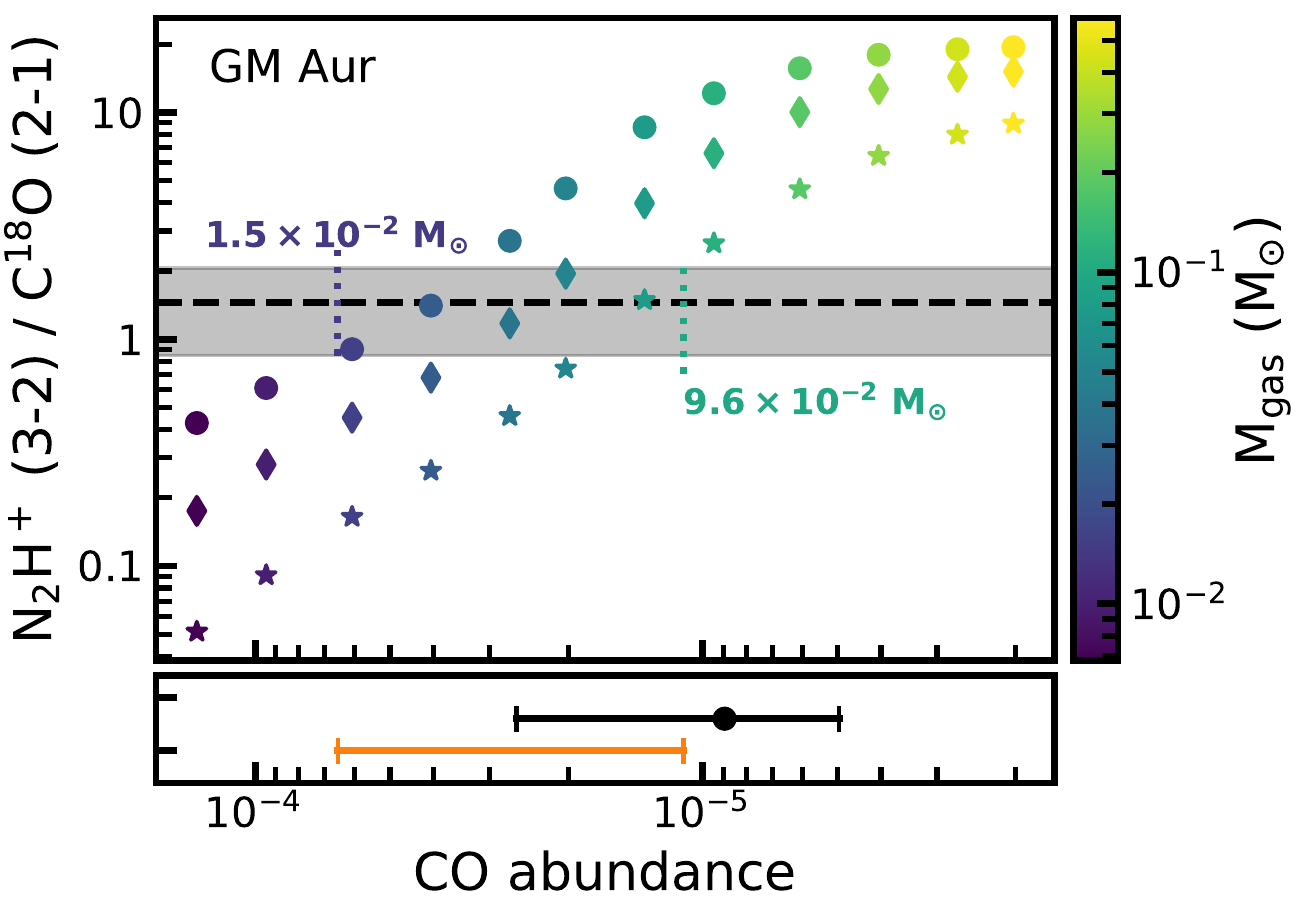}
    \end{minipage}
    \caption{\label{fig: Flux ratio} \nhp\,(3-2)/\cyo\,(2-1) flux ratio versus the CO abundance. Points denote individual models, where the shape shows the assumed cosmic ray ionization rate $(\zet)$. The colors show the gas mass of the disk. The observed \nhp\,(3-2)/\cyo\,(2-1) ratio is shown as a black dashed line, with the gray shaded region showing the $3\sigma$ uncertainty. For reference we included the lowest and highest disk mass that reproduce these observations, assuming $10^{-19}\ \mathrm{s}^{-1}\leq\zet\leq10^{-17}\ \mathrm{s}^{-1}$. The smaller panels under each main panel show the CO abundance ranges that reproduce the observations. Here the orange bar corresponds to \nhp\,(3-2) and \cyo\,(2-1), with $10^{-19}\ \mathrm{s}^{-1}\leq\zet\leq10^{-17}\ \mathrm{s}^{-1}$. The black bar instead shows the CO abundance constrained by HD 1-0 and \cyo\,(2-1).   }
\end{figure}

Figure \ref{fig: Flux ratio} shows the \nhp\ (3-2)/\cyo\ (2-1) integrated flux ratio versus the CO abundance for the three disks. Note that each of the models reproduces the \cyo\ integrated flux within $3\sigma$. The flux ratio clearly increases as the CO abundance decreases. This trend is similar to the one found by \cite{Anderson2019} (see their Figures 5 and 9). For GM Aur the trend flattens at the highest disk masses where both lines start to become optically thick. The figure also shows that lowering \zet\ reduces the flux ratio. A lower \zet\ equals less ionization and therefore less \nhp\ and a lower \nhp\ 3-2 flux. The \cyo\ flux is not affected, but we should note that our chemical network does not include the chemical conversion of CO into more complex species, where \zet\ plays a large role (see \cite{Schwarz2018,Schwarz2019,Bosman2018b}).
By comparing our models to the observations shown in gray, we can derive the global CO abundance of the disk.

Both TW Hya and GM Aur require a CO abundance $\leq10^{-4}$ to match the \nhp/\cyo\ line ratio. Specifically, TW Hya requires $\abu \leq 1.3\times10^{-5}$ and GM Aur requires $\abu \leq 6.5\times10^{-5}$. 
For DM Tau the CO abundance does not have to be lowered to match the line ratio. Rather the models show that a lower cosmic ionization rate $\sim\,5\times10^{-18}\ \mathrm{s}^{-1}$ is needed to reproduce the line ratio with a CO abundance of $10^{-4}$.
The lower bound on the CO abundance depends on the cosmic ionization rate. If we assume a lower bound of $\zet = 10^{-19}\ \mathrm{s}^{-1}$ in disks (see, e.g. \citealt{Cleeves2015}), we can constrain the CO abundances to $3.5\times10^{-7} \leq \abu \leq 1.3\times10^{-5}$ for TW Hya, $1.1\times10^{-5}\leq \abu \leq 6.5\times10^{-5}$ for GM Aur and $2.3\times10^{-5}\leq \abu \leq 1.3\times10^{-4}$ for DM Tau. 
As shown here, a two order of magnitude uncertainty in \zet\ results in a factor of $5-37$ uncertainty in the CO abundance and thus the gas disk mass. Constraining the cosmic-ray ionization rate in disks using observations of, for example, H$^{13}$CO$^+$ and N$_2$D$^+$ will reduce the uncertainty on the gas disk mass (see, e.g., \citealt{Cleeves2015,AikawaMAPS2021}).

Using these CO abundances we can now correct the CO-based gas mass for the underabundance of CO and obtain the true gas disk mass. This yields $4.6\times10^{-3}\ \msun \leq \mdisk \leq 1.1\times10^{-1}\ \msun$ for TW Hya, $1.5\times10^{-2}\ \msun\leq \mdisk \leq 9.6\times10^{-2}\ \msun$ for GM Aur and $3.1\times10^{-2}\ \msun \leq \mdisk \leq 9.6\times10^{-2}\msun$ for DM Tau. By including \nhp\ we have increased the accuracy of the measured gas disk mass by a factor 5-10 compared to CO-based gas masses, by correcting for the fact that CO is underabundant in the gas by a factor of 5-10 in our disks. As mentioned above, the precision of the gas mass measurement can be improved by constraining the cosmic ray ionization rate. 
Thus, combining \nhp\ and \cyo\ allows us to break the degeneracy between gas disk mass and CO abundance shown in Figure \ref{fig: C18O score} and gives us a recipe for measuring gas disk masses.

It is interesting to compare our new gas disk masses to those measured independently from HD 1-0 integrated line fluxes. To be consistent with the disk structures assumed in the rest of this work, we opt to use our models to measure $M_{\rm gas}^{HD}$ rather than using literature values (see Table \ref{tab:source_info}). Figure \ref{fig: HD gas masses} in Appendix \ref{app: HD gas masses} shows how $M_{\rm gas}^{HD}$ is derived from the HD 1-0 integrated flux. Uncertainties on the masses are derived by propagating the $3\sigma$ uncertainties of the line flux. The HD-based based disk masses for TW Hya and GM Aur are consistent with literature values (see Table \ref{tab:source_info}). Only for DM Tau do we find a substantially higher disk mass than previously found, due to our model being more extended, and thus having lower average gas temperature, than the model in \cite{McClure2016}.  
As HD predominantly emits from warm gas, it should be noted that the accuracy of HD as a gas mass tracer depends on how well the temperature structure of the disk is known (see e.g. \citealt{Trapman2017,Calahan2021}). If the disks are significantly warmer than our models we would overestimate their gas mass. This is unlikely to be a large effect however as the disk structures used here reproduce several optically thick CO lines which trace the gas temperature.

The disk mass obtained for DM Tau warrants some further discussion. Based on the HD 1-0 flux and uncertainty, we find a maximum disk mass of $\mgas = 0.55\ \msun$, which would make the disk approximately as massive as the star. Such a high disk mass is unphysical, as disk would become gravitationally unstable before it could reach this disk mass. To find a more physical maximum disk mass, we calculate the Toomre Q parameter \cite{Toomre1964} using the surface density and midplane temperature profiles from our models. For a gas disk mass $\mgas \approx 0.2\ \msun$ we find $Q \lesssim 1.7$, which is the approximate threshold where instabilities can start to develop in numerical simulations of disks (e.g. \citealt{Helled2014}). For the rest of this work we will use $\mgas = 0.2\ \msun$ as a more realistic maximum disk mass for DM Tau.  

\begin{figure*}
    \centering
    \includegraphics[width=0.8\textwidth]{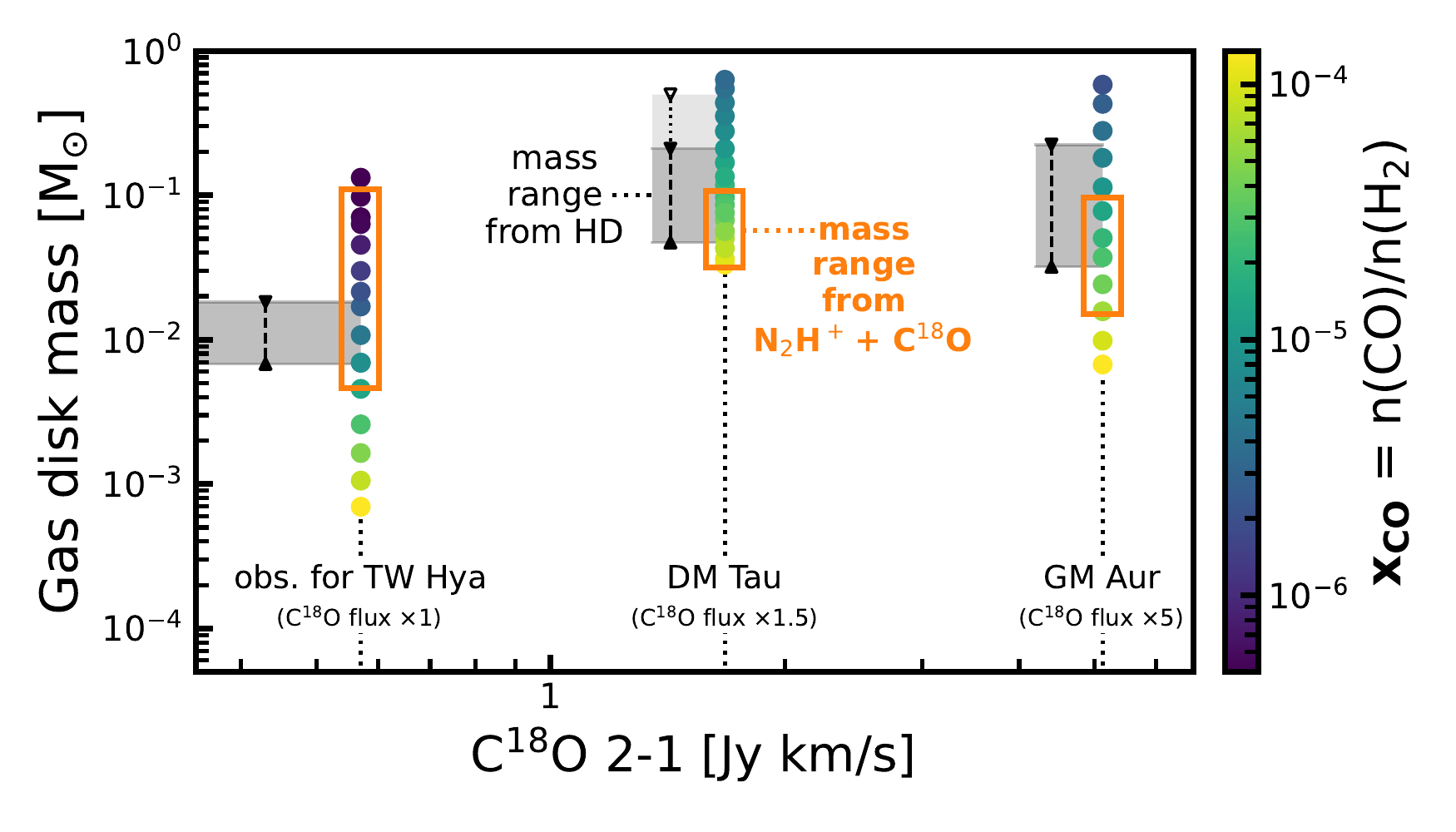}
    \caption{\label{fig: comparing constraints} 
    Comparison of the mass ranges obtained from HD 1-0 (gray shaded region) and the combination of \nhp\ and \cyo\ (orange box). Note that for \nhp\ and \cyo\ we have assumed $10^{-19}\ \mathrm{s}^{-1}\leq\zet\leq10^{-17}\ \mathrm{s}^{-1}$. Points show individual models, with their color showing the CO abundance. For DM Tau, the transition from dark to light gray shows where $\mdisk = 0.2\ \msun$. For a higher disk mass there should be signs of gravitational instability in the disk, which have not been seen in observations 
    (see Section \ref{sec: results} for more details).}
\end{figure*}

Figure \ref{fig: comparing constraints} compares the gas mass obtained from combining \nhp\ and \cyo\ to the gas mass measured from HD 1-0 (see Figure \ref{fig: HD gas masses}). Lining up the two mass ranges of each of the three sources, we find good agreement. For DM Tau and GM Aur the \nhp\ + \cyo\ gas masses are overall slightly lower compared to the HD gas mass. The lower end of the \nhp\ + \cyo\ gas mass range corresponds to the highest assumed \zet, suggesting that these disks have a lower cosmic ray ionization rate (see Section \ref{sec: constraining the cosmic ray ionization rate} for a further discussion on this topic). 
The uncertainty on the \nhp\ + \cyo\ gas mass for DM Tau and GM Aur is similar to the uncertainty of the HD-based gas mass measured from their low signal-to-noise HD 1-0 detections.
For TW Hya, where HD was detected with comparatively high signal-to-noise, the resulting $M_{\rm gas}^{HD}$ range is more narrow and matches the gas mass derived from \nhp\ and \cyo\ if a high cosmic ray ionization rate is used.

While based on only a small sample of disks, Figure \ref{fig: comparing constraints} shows that combination of \nhp\ and \cyo\ integrated line fluxes is a promising new recipe for measuring gas masses of protoplanetary disks

\section{Discussion}
\label{sec: discussion}

\subsection{Constraining the cosmic ray ionization rate}
\label{sec: constraining the cosmic ray ionization rate}
As previously discussed, the gas disk mass derived from \nhp\ and \cyo\ depends on the assumed \zet. In our comparison between different gas disk mass estimates we have taken the dependence on \zet\ as an addition uncertainty on the CO abundance estimated from \nhp. However, we can also examine the overlap between the gas disk mass from HD and one from \nhp\ and \cyo\ to put constraints on \zet. 
Taking the gas disk mass derived from the HD observations, we determine the range of \zet\ for which our models are able to reproduce the \nhp\ 3-2 integrated flux.

We find a clear dichotomy in cosmic ray ionization rates. 
One disk, TW Hya, is most consistent with a relatively high cosmic ray ionization rate, $5\times10^{-18} \leq \zet \leq 1\times10^{-17}\ \mathrm{s}^{-1}$, while the other two disks 
are more consistent with low rates, $\zet \leq 1\times10^{-18}\ \mathrm{s}^{-1}$ for DM Tau and $\zet \leq 5\times10^{-19}\ \mathrm{s}^{-1}$ for GM Aur, respectively. 
The low \zet\ for DM Tau and GM Aur are consistent with earlier findings (see, e.g. \citealt{Cleeves2015}). In particular, \cite{AikawaMAPS2021} found a low cosmic ray ionization $(\leq10^{-18}\ \mathrm{s}^{-1})$ for GM Aur based on deep observations of \nhp\ and N$_2$D$^+$. 
Finding a high \zet\ for TW Hya is more puzzling. In an earlier study, \cite{Cleeves2015} modeled \nhp, HCO$^+$ and H$^{13}$CO$^+$, in combination with CO isotopologues and HD, observations of TW Hya, finding that the observations are best reproduced by a moderately hard X-ray spectrum and a low cosmic ray ionization rate $(\zet \leq 10^{-19}\mathrm{s}^{-1})$. In their models, a high \zet\ such as the one found in this work results in a peak of the \nhp\ 4-3 emission at $\sim\,130$ au, much further out than the observed location at $\sim\,45$ au (see \citealt{Qi2013}). Interestingly, the peak of \nhp\ 4-3 radial intensity profile in our TW Hya model with $\zet = 10^{-17}\ \mathrm{s}^{-1}$ lies further inward and matches the observed peak location (cf. \citealt{vtHoff2017}). 

There are three key differences between the models in \cite{Cleeves2015} and the models presented in this work. The first is the assumed surface density structure. Our TW Hya models have a characteristic radius of $R_c =35$ au, whereas the models in \cite{Cleeves2015} use $R_c = 150$ au and truncate the disk at $200$ au. However, \cite{Cleeves2015} showed that this difference in surface density has minimal impact on the \nhp\ column densities (see their appendix C). A similar test where we change the surface density in our models confirms that this cannot explain the difference in \nhp\ 4-3 intensity profiles. 
The second is the chemical network.
In this work we use a simplified \nhp\ chemical network that only includes its dominant formation and destruction pathways. This suggests that there are one or more reactions in network used in \citealt{Cleeves2015} that significantly affect the \nhp\ chemistry that are not included in our network. One such reaction could be the non-thermal desorption of N$_2$ and CO. However, \citealt{vtHoff2017} showed that an increased desorption rate has a negligible effect on the \nhp\ 4-3 emission profile (see their Appendix D).
Finally, we only include ionization through cosmic rays, which is the dominant source for ionization close to the disk midplane. \cite{Cleeves2015} also include X-ray ionization, which could affect \nhp\ abundance higher up in the disk.
Reconciling the two models requires a full comparison between the two approaches, which is beyond the scope of this paper. We do note that the \cite{Cleeves2015} model with $\zet \approx 10^{-18}\ \mathrm{s}^{-1}$ combined with a softer X-ray spectrum also reproduces the observed \nhp\ 4-3 emission, which is more similar to the value found in this work.

\subsection{Caveats}
\label{sec: caveats}
Here we briefly discuss some of the caveats and uncertainties that could effect our results. A detailed investigation of these caveats is beyond the scope of this Letter and will be reserved for a forthcoming paper. 

The gas and dust density structures used in this work fit the observations, but they are not unique in doing so (see, e.g. \citealt{Calahan2021}). The assumed disk structure could thus affect our results. However, our comparison with the gas density structure of the \cite{Cleeves2015} model discussed in Section \ref{sec: constraining the cosmic ray ionization rate} suggests that the effect on the \nhp\ flux is small, $\leq 50\%$.
Note that our disk structure also does not include the gaps and rings that have been found in the continuum emission, except for the inner cavity. Increased X-ray ionization in such a gap could increase the \nhp\ abundance, but the higher CO abundance due to the increased temperature could decrease the \nhp\ again, making it unclear what the net effect on the \nhp\ abundance will be (e.g. \citealt{KimTurner2020,Alarcon2020}).

We have also assumed that N$_2$ is the dominant nitrogen carrier in the gas and have not varied its abundance in our analysis. While this is likely a good assumption, it has not been conclusively shown observationally (e.g., \citealt{Salinas2016,Cleeves2018}). 
If N$_2$ is not the dominant nitrogen carrier both the \nhp\ abundance and flux would be lower. To still reproduce the observed \nhp\ 3-2 flux either our derived CO abundances would need to be lower, implying a larger disk mass, or cosmic ray ionization rates would need to be higher. Tests show that lowering N$_2$ abundance by a factor of ten, equivalent to assuming that 10\% of the volatile nitrogen is in N$_2$, decreases the \nhp\ flux by a similar amount as reducing \zet\ by two orders of magnitude.
The assumed binding energies of N$_2$ and CO could also affect the \nhp\ abundance, but \citealt{vtHoff2017} showed that their effect on the \nhp\ flux is minimal.

\section{Conclusions}
\label{sec: conclusions}

The gas mass of protoplanetary disks remains an important yet elusive quantity.
In this work we present a novel recipe for measuring the gas mass, where we use \nhp\ to observationally measure the global CO abundance in disks, the crucial parameter for deriving gas masses from \cyo\ observations.
We test this method for the three disks, TW Hya, DM Tau and GM Aur, where we compare the resulting gas mass to the independently measured gas masses obtained from fitting the HD 1-0 flux.
We summarize our findings below:

\begin{itemize}
    \item The \nhp($J=3-2$)/\cyo($J=2-1$) line ratio from our models scales with the CO-to-H$_2$ ratio, confirming earlier findings by \citealt{Anderson2019}. This shows that the combination of these two lines can be used to observationally constrain the CO-to-H$_2$ ratio in disks, and hence their gas mass.
    \item Using the combination of \nhp\ and \cyo, we measure $4.6\times10^{-3}\ \msun \leq \mdisk \leq 1.1\times10^{-1}\ \msun$ for TW Hya, $1.5\times10^{-2}\ \msun\leq \mdisk \leq 9.6\times10^{-2}\ \msun$ for GM Aur and $3.1\times10^{-2}\ \msun \leq \mdisk \leq 9.6\times10^{-2}\msun$ for DM Tau, respectively. Including \nhp\ increases the accuracy of the measured gas mass by a factor of 5-10, by correcting for the underabundance of gaseous CO.
    These gas masses agree with the disk gas masses measured from the HD 1-0 line flux to within their respective uncertainties for each of our three sources. 
    \item The cosmic ray ionization rate \zet\ is the main uncertainty on how well the CO-to-H$_2$ ratio, and thus the gas mass can be measured, as a lower \zet\ directly decreases the \nhp\ 3-2 flux. For $10^{-19}\ \mathrm{s}^{-1} \leq \zet \leq 10^{-17}\ \mathrm{s}^{-1}$, the uncertainty in the gas mass is a factor of $5-37$. 
    Further observations of ionization tracers such as H$^{13}$CO$^+$ and N$_2$D$^+$ that constrain the cosmic ray ionization rate in disks can help reduce the uncertainty in the measured gas disk mass.
\end{itemize}

The agreement between the disk gas mass measured from \nhp\ and \cyo\ and the independently measured gas mass from HD shows that combining \nhp\ and \cyo\ is a promising new way of measuring the total gas reservoir of planet-forming disks.

\begin{acknowledgements}
The authors thank the referee for their constructive comments and Ilse Cleeves for sharing the \nhp\ observations of TW Hya.
The authors would also like to thank Ilse Cleeves, John Carpenter, Lucas Cieza, Laura Perez and Paola Pinilla for the useful discussions that let to this work.
L.T. and K. Z. acknowledges the support of the Office of the Vice Chancellor for Research and Graduate Education at the University of Wisconsin – Madison with funding from the Wisconsin Alumni Research Foundation.
M.L.R.H acknowledges support from the Michigan Society of Fellows.
This paper makes use of the following ALMA data:
ADS/JAO.ALMA\#2016.1.00311.S. ALMA is a partnership of ESO (representing its member states), NSF (USA) and NINS (Japan), together with NRC (Canada), MOST and ASIAA (Taiwan), and KASI (Republic of Korea), in cooperation with the Republic of Chile. The Joint ALMA Observatory is operated by ESO, AUI/NRAO and NAOJ.
All figures were generated with the \texttt{PYTHON}-based package \texttt{MATPLOTLIB} \citep{Hunter2007}. This research made use of NumPy \citep{Harris2020} and Astropy,\footnote{http://www.astropy.org} a community-developed core Python package for Astronomy \citep{astropy:2013, astropy:2018}.
\end{acknowledgements}

\bibliographystyle{aasjournal}
\bibliography{references}

\begin{thebibliography}{}
\expandafter\ifx\csname natexlab\endcsname\relax\def\natexlab#1{#1}\fi
\providecommand{\url}[1]{\href{#1}{#1}}
\providecommand{\dodoi}[1]{doi:~\href{http://doi.org/#1}{\nolinkurl{#1}}}
\providecommand{\doeprint}[1]{\href{http://ascl.net/#1}{\nolinkurl{http://ascl.net/#1}}}
\providecommand{\doarXiv}[1]{\href{https://arxiv.org/abs/#1}{\nolinkurl{https://arxiv.org/abs/#1}}}

\bibitem[{{Aikawa} {et~al.}(2015){Aikawa}, {Furuya}, {Nomura}, \&
  {Qi}}]{Aikawa2015}
{Aikawa}, Y., {Furuya}, K., {Nomura}, H., \& {Qi}, C. 2015, \apj, 807, 120,
  \dodoi{10.1088/0004-637X/807/2/120}

\bibitem[{Aikawa {et~al.}(1997)Aikawa, Umebayashi, Nakano, \&
  Miyama}]{Aikawa1997}
Aikawa, Y., Umebayashi, T., Nakano, T., \& Miyama, S.~M. 1997, \apjl, 486, L51

\bibitem[{{Aikawa} {et~al.}(2021){Aikawa}, {Cataldi}, {Yamato}, {Zhang},
  {Booth}, {Furuya}, {Andrews}, {Bae}, {Bergin}, {Bergner}, {Bosman},
  {Cleeves}, {Czekala}, {Guzm{\'a}n}, {Huang}, {Ilee}, {Law}, {Le Gal},
  {Loomis}, {M{\'e}nard}, {Nomura}, {{\"O}berg}, {Qi}, {Schwarz}, {Teague},
  {Tsukagoshi}, {Walsh}, \& {Wilner}}]{AikawaMAPS2021}
{Aikawa}, Y., {Cataldi}, G., {Yamato}, Y., {et~al.} 2021, \apjs, 257, 13,
  \dodoi{10.3847/1538-4365/ac143c}

\bibitem[{{Alarc{\'o}n} {et~al.}(2020){Alarc{\'o}n}, {Teague}, {Zhang},
  {Bergin}, \& {Barraza-Alfaro}}]{Alarcon2020}
{Alarc{\'o}n}, F., {Teague}, R., {Zhang}, K., {Bergin}, E.~A., \&
  {Barraza-Alfaro}, M. 2020, \apj, 905, 68, \dodoi{10.3847/1538-4357/abc1d6}

\bibitem[{{Anderson} {et~al.}(2019){Anderson}, {Blake}, {Bergin}, {Zhang},
  {Carpenter}, {Schwarz}, {Huang}, \& {{\"O}berg}}]{Anderson2019}
{Anderson}, D.~E., {Blake}, G.~A., {Bergin}, E.~A., {et~al.} 2019, \apj, 881,
  127, \dodoi{10.3847/1538-4357/ab2cb5}

\bibitem[{Andrews {et~al.}(2011)Andrews, Wilner, Hughes, Qi, Rosenfeld,
  {\"O}berg, Birnstiel, Espaillat, Cieza, Williams, {et~al.}}]{Andrews2012}
Andrews, S.~M., Wilner, D.~J., Hughes, A., {et~al.} 2011, \apj, 744, 162

\bibitem[{Ansdell {et~al.}(2016)Ansdell, Williams, van~der Marel, Carpenter,
  Guidi, Hogerheijde, Mathews, Manara, Miotello, Natta, {et~al.}}]{ansdell2016}
Ansdell, M., Williams, J.~P., van~der Marel, N., {et~al.} 2016, \apj, 828, 46

\bibitem[{{Astropy Collaboration} {et~al.}(2013){Astropy Collaboration},
  {Robitaille}, {Tollerud}, {Greenfield}, {Droettboom}, {Bray}, {Aldcroft},
  {Davis}, {Ginsburg}, {Price-Whelan}, {Kerzendorf}, {Conley}, {Crighton},
  {Barbary}, {Muna}, {Ferguson}, {Grollier}, {Parikh}, {Nair}, {Unther},
  {Deil}, {Woillez}, {Conseil}, {Kramer}, {Turner}, {Singer}, {Fox}, {Weaver},
  {Zabalza}, {Edwards}, {Azalee Bostroem}, {Burke}, {Casey}, {Crawford},
  {Dencheva}, {Ely}, {Jenness}, {Labrie}, {Lim}, {Pierfederici}, {Pontzen},
  {Ptak}, {Refsdal}, {Servillat}, \& {Streicher}}]{astropy:2013}
{Astropy Collaboration}, {Robitaille}, T.~P., {Tollerud}, E.~J., {et~al.} 2013,
  \aap, 558, A33, \dodoi{10.1051/0004-6361/201322068}

\bibitem[{{Astropy Collaboration} {et~al.}(2018){Astropy Collaboration},
  {Price-Whelan}, {Sip{\H{o}}cz}, {G{\"u}nther}, {Lim}, {Crawford}, {Conseil},
  {Shupe}, {Craig}, {Dencheva}, {Ginsburg}, {Vand erPlas}, {Bradley},
  {P{\'e}rez-Su{\'a}rez}, {de Val-Borro}, {Aldcroft}, {Cruz}, {Robitaille},
  {Tollerud}, {Ardelean}, {Babej}, {Bach}, {Bachetti}, {Bakanov}, {Bamford},
  {Barentsen}, {Barmby}, {Baumbach}, {Berry}, {Biscani}, {Boquien}, {Bostroem},
  {Bouma}, {Brammer}, {Bray}, {Breytenbach}, {Buddelmeijer}, {Burke},
  {Calderone}, {Cano Rodr{\'\i}guez}, {Cara}, {Cardoso}, {Cheedella}, {Copin},
  {Corrales}, {Crichton}, {D'Avella}, {Deil}, {Depagne}, {Dietrich}, {Donath},
  {Droettboom}, {Earl}, {Erben}, {Fabbro}, {Ferreira}, {Finethy}, {Fox},
  {Garrison}, {Gibbons}, {Goldstein}, {Gommers}, {Greco}, {Greenfield},
  {Groener}, {Grollier}, {Hagen}, {Hirst}, {Homeier}, {Horton}, {Hosseinzadeh},
  {Hu}, {Hunkeler}, {Ivezi{\'c}}, {Jain}, {Jenness}, {Kanarek}, {Kendrew},
  {Kern}, {Kerzendorf}, {Khvalko}, {King}, {Kirkby}, {Kulkarni}, {Kumar},
  {Lee}, {Lenz}, {Littlefair}, {Ma}, {Macleod}, {Mastropietro}, {McCully},
  {Montagnac}, {Morris}, {Mueller}, {Mumford}, {Muna}, {Murphy}, {Nelson},
  {Nguyen}, {Ninan}, {N{\"o}the}, {Ogaz}, {Oh}, {Parejko}, {Parley}, {Pascual},
  {Patil}, {Patil}, {Plunkett}, {Prochaska}, {Rastogi}, {Reddy Janga},
  {Sabater}, {Sakurikar}, {Seifert}, {Sherbert}, {Sherwood-Taylor}, {Shih},
  {Sick}, {Silbiger}, {Singanamalla}, {Singer}, {Sladen}, {Sooley},
  {Sornarajah}, {Streicher}, {Teuben}, {Thomas}, {Tremblay}, {Turner},
  {Terr{\'o}n}, {van Kerkwijk}, {de la Vega}, {Watkins}, {Weaver}, {Whitmore},
  {Woillez}, {Zabalza}, \& {Astropy Contributors}}]{astropy:2018}
{Astropy Collaboration}, {Price-Whelan}, A.~M., {Sip{\H{o}}cz}, B.~M., {et~al.}
  2018, \aj, 156, 123, \dodoi{10.3847/1538-3881/aabc4f}

\bibitem[{Bergin {et~al.}(2010)Bergin, Hogerheijde, Brinch, Fogel,
  Y{\i}ld{\i}z, Kristensen, Van~Dishoeck, Bell, Blake, Cernicharo,
  {et~al.}}]{Bergin2010}
Bergin, E., Hogerheijde, M., Brinch, C., {et~al.} 2010, \aap, 521, L33

\bibitem[{{Bergin} {et~al.}(2016){Bergin}, {Du}, {Cleeves}, {Blake}, {Schwarz},
  {Visser}, \& {Zhang}}]{Bergin2016}
{Bergin}, E.~A., {Du}, F., {Cleeves}, L.~I., {et~al.} 2016, \apj, 831, 101,
  \dodoi{10.3847/0004-637X/831/1/101}

\bibitem[{{Bergin} {et~al.}(2013){Bergin}, {Cleeves}, {Gorti}, {Zhang},
  {Blake}, {Green}, {Andrews}, {Evans}, {Henning}, {{\"O}berg}, {Pontoppidan},
  {Qi}, {Salyk}, \& {van Dishoeck}}]{Bergin2013}
{Bergin}, E.~A., {Cleeves}, L.~I., {Gorti}, U., {et~al.} 2013, \nat, 493, 644,
  \dodoi{10.1038/nature11805}

\bibitem[{{Bergner} {et~al.}(2019){Bergner}, {{\"O}berg}, {Bergin}, {Loomis},
  {Pegues}, \& {Qi}}]{Bergner2019}
{Bergner}, J.~B., {{\"O}berg}, K.~I., {Bergin}, E.~A., {et~al.} 2019, \apj,
  876, 25, \dodoi{10.3847/1538-4357/ab141e}

\bibitem[{Birnstiel {et~al.}(2012)Birnstiel, Klahr, \&
  Ercolano}]{Birnstiel2012}
Birnstiel, T., Klahr, H., \& Ercolano, B. 2012, \aap, 539, A148

\bibitem[{{Booth} {et~al.}(2019){Booth}, {Walsh}, {Ilee}, {Notsu}, {Qi},
  {Nomura}, \& {Akiyama}}]{Booth2019}
{Booth}, A.~S., {Walsh}, C., {Ilee}, J.~D., {et~al.} 2019, \apjl, 882, L31,
  \dodoi{10.3847/2041-8213/ab3645}

\bibitem[{{Bosman} {et~al.}(2018){Bosman}, {Walsh}, \& {van
  Dishoeck}}]{Bosman2018b}
{Bosman}, A.~D., {Walsh}, C., \& {van Dishoeck}, E.~F. 2018, \aap, 618, A182,
  \dodoi{10.1051/0004-6361/201833497}

\bibitem[{{Brickhouse} {et~al.}(2010){Brickhouse}, {Cranmer}, {Dupree}, {Luna},
  \& {Wolk}}]{Brickhouse2010}
{Brickhouse}, N.~S., {Cranmer}, S.~R., {Dupree}, A.~K., {Luna}, G.~J.~M., \&
  {Wolk}, S. 2010, \apj, 710, 1835, \dodoi{10.1088/0004-637X/710/2/1835}

\bibitem[{{Bruderer}(2013)}]{Bruderer2013}
{Bruderer}, S. 2013, \aap, 559, A46, \dodoi{10.1051/0004-6361/201321171}

\bibitem[{{Bruderer} {et~al.}(2012){Bruderer}, {van Dishoeck}, {Doty}, \&
  {Herczeg}}]{Bruderer2012}
{Bruderer}, S., {van Dishoeck}, E.~F., {Doty}, S.~D., \& {Herczeg}, G.~J. 2012,
  \aap, 541, A91, \dodoi{10.1051/0004-6361/201118218}

\bibitem[{{Calahan} {et~al.}(2021){Calahan}, {Bergin}, {Zhang}, {Teague},
  {Cleeves}, {Bergner}, {Blake}, {Cazzoletti}, {Guzm{\'a}n}, {Hogerheijde},
  {Huang}, {Kama}, {Loomis}, {{\"O}berg}, {Qi}, {van Dishoeck}, {Terwisscha van
  Scheltinga}, {Walsh}, \& {Wilner}}]{Calahan2021}
{Calahan}, J.~K., {Bergin}, E., {Zhang}, K., {et~al.} 2021, \apj, 908, 8,
  \dodoi{10.3847/1538-4357/abd255}

\bibitem[{{Cleeves} {et~al.}(2015){Cleeves}, {Bergin}, {Qi}, {Adams}, \&
  {{\"O}berg}}]{Cleeves2015}
{Cleeves}, L.~I., {Bergin}, E.~A., {Qi}, C., {Adams}, F.~C., \& {{\"O}berg},
  K.~I. 2015, \apj, 799, 204, \dodoi{10.1088/0004-637X/799/2/204}

\bibitem[{{Cleeves} {et~al.}(2018){Cleeves}, {{\"O}berg}, {Wilner}, {Huang},
  {Loomis}, {Andrews}, \& {Guzman}}]{Cleeves2018}
{Cleeves}, L.~I., {{\"O}berg}, K.~I., {Wilner}, D.~J., {et~al.} 2018, \apj,
  865, 155, \dodoi{10.3847/1538-4357/aade96}

\bibitem[{{Dionatos} {et~al.}(2019){Dionatos}, {Woitke}, {G{\"u}del},
  {Degroote}, {Liebhart}, {Anthonioz}, {Antonellini}, {Baldovin-Saavedra},
  {Carmona}, {Dominik}, {Greaves}, {Ilee}, {Kamp}, {M{\'e}nard}, {Min},
  {Pinte}, {Rab}, {Rigon}, {Thi}, \& {Waters}}]{Dionatos2019}
{Dionatos}, O., {Woitke}, P., {G{\"u}del}, M., {et~al.} 2019, \aap, 625, A66,
  \dodoi{10.1051/0004-6361/201832860}

\bibitem[{{Favre} {et~al.}(2013){Favre}, {Cleeves}, {Bergin}, {Qi}, \&
  {Blake}}]{Favre2013}
{Favre}, C., {Cleeves}, L.~I., {Bergin}, E.~A., {Qi}, C., \& {Blake}, G.~A.
  2013, \apjl, 776, L38, \dodoi{10.1088/2041-8205/776/2/L38}

\bibitem[{{Furuya} \& {Aikawa}(2014)}]{FuruyaAikawa2014}
{Furuya}, K., \& {Aikawa}, Y. 2014, \apj, 790, 97,
  \dodoi{10.1088/0004-637X/790/2/97}

\bibitem[{{Gaia Collaboration} {et~al.}(2018){Gaia Collaboration}, {Brown},
  {Vallenari}, {Prusti}, {de Bruijne}, {Babusiaux}, {Bailer-Jones}, {Biermann},
  {Evans}, {Eyer}, {Jansen}, {Jordi}, {Klioner}, {Lammers}, {Lindegren},
  {Luri}, {Mignard}, {Panem}, {Pourbaix}, {Randich}, {Sartoretti}, {Siddiqui},
  {Soubiran}, {van Leeuwen}, {Walton}, {Arenou}, {Bastian}, {Cropper},
  {Drimmel}, {Katz}, {Lattanzi}, {Bakker}, {Cacciari}, {Casta{\~n}eda},
  {Chaoul}, {Cheek}, {De Angeli}, {Fabricius}, {Guerra}, {Holl}, {Masana},
  {Messineo}, {Mowlavi}, {Nienartowicz}, {Panuzzo}, {Portell}, {Riello},
  {Seabroke}, {Tanga}, {Th{\'e}venin}, {Gracia-Abril}, {Comoretto},
  {Garcia-Reinaldos}, {Teyssier}, {Altmann}, {Andrae}, {Audard},
  {Bellas-Velidis}, {Benson}, {Berthier}, {Blomme}, {Burgess}, {Busso},
  {Carry}, {Cellino}, {Clementini}, {Clotet}, {Creevey}, {Davidson}, {De
  Ridder}, {Delchambre}, {Dell'Oro}, {Ducourant},
  {Fern{\'a}ndez-Hern{\'a}ndez}, {Fouesneau}, {Fr{\'e}mat}, {Galluccio},
  {Garc{\'\i}a-Torres}, {Gonz{\'a}lez-N{\'u}{\~n}ez}, {Gonz{\'a}lez-Vidal},
  {Gosset}, {Guy}, {Halbwachs}, {Hambly}, {Harrison}, {Hern{\'a}ndez},
  {Hestroffer}, {Hodgkin}, {Hutton}, {Jasniewicz}, {Jean-Antoine-Piccolo},
  {Jordan}, {Korn}, {Krone-Martins}, {Lanzafame}, {Lebzelter}, {L{\"o}ffler},
  {Manteiga}, {Marrese}, {Mart{\'\i}n-Fleitas}, {Moitinho}, {Mora}, {Muinonen},
  {Osinde}, {Pancino}, {Pauwels}, {Petit}, {Recio-Blanco}, {Richards},
  {Rimoldini}, {Robin}, {Sarro}, {Siopis}, {Smith}, {Sozzetti}, {S{\"u}veges},
  {Torra}, {van Reeven}, {Abbas}, {Abreu Aramburu}, {Accart}, {Aerts},
  {Altavilla}, {{\'A}lvarez}, {Alvarez}, {Alves}, {Anderson}, {Andrei},
  {Anglada Varela}, {Antiche}, {Antoja}, {Arcay}, {Astraatmadja}, {Bach},
  {Baker}, {Balaguer-N{\'u}{\~n}ez}, {Balm}, {Barache}, {Barata}, {Barbato},
  {Barblan}, {Barklem}, {Barrado}, {Barros}, {Barstow}, {Bartholom{\'e}
  Mu{\~n}oz}, {Bassilana}, {Becciani}, {Bellazzini}, {Berihuete}, {Bertone},
  {Bianchi}, {Bienaym{\'e}}, {Blanco-Cuaresma}, {Boch}, {Boeche}, {Bombrun},
  {Borrachero}, {Bossini}, {Bouquillon}, {Bourda}, {Bragaglia}, {Bramante},
  {Breddels}, {Bressan}, {Brouillet}, {Br{\"u}semeister}, {Brugaletta},
  {Bucciarelli}, {Burlacu}, {Busonero}, {Butkevich}, {Buzzi}, {Caffau},
  {Cancelliere}, {Cannizzaro}, {Cantat-Gaudin}, {Carballo}, {Carlucci},
  {Carrasco}, {Casamiquela}, {Castellani}, {Castro-Ginard}, {Charlot},
  {Chemin}, {Chiavassa}, {Cocozza}, {Costigan}, {Cowell}, {Crifo}, {Crosta},
  {Crowley}, {Cuypers}, {Dafonte}, {Damerdji}, {Dapergolas}, {David}, {David},
  {de Laverny}, {De Luise}, {De March}, {de Martino}, {de Souza}, {de Torres},
  {Debosscher}, {del Pozo}, {Delbo}, {Delgado}, {Delgado}, {Di Matteo},
  {Diakite}, {Diener}, {Distefano}, {Dolding}, {Drazinos}, {Dur{\'a}n},
  {Edvardsson}, {Enke}, {Eriksson}, {Esquej}, {Eynard Bontemps}, {Fabre},
  {Fabrizio}, {Faigler}, {Falc{\~a}o}, {Farr{\`a}s Casas}, {Federici},
  {Fedorets}, {Fernique}, {Figueras}, {Filippi}, {Findeisen}, {Fonti},
  {Fraile}, {Fraser}, {Fr{\'e}zouls}, {Gai}, {Galleti}, {Garabato},
  {Garc{\'\i}a-Sedano}, {Garofalo}, {Garralda}, {Gavel}, {Gavras}, {Gerssen},
  {Geyer}, {Giacobbe}, {Gilmore}, {Girona}, {Giuffrida}, {Glass}, {Gomes},
  {Granvik}, {Gueguen}, {Guerrier}, {Guiraud}, {Guti{\'e}rrez-S{\'a}nchez},
  {Haigron}, {Hatzidimitriou}, {Hauser}, {Haywood}, {Heiter}, {Helmi}, {Heu},
  {Hilger}, {Hobbs}, {Hofmann}, {Holland}, {Huckle}, {Hypki}, {Icardi},
  {Jan{\ss}en}, {Jevardat de Fombelle}, {Jonker}, {Juh{\'a}sz}, {Julbe},
  {Karampelas}, {Kewley}, {Klar}, {Kochoska}, {Kohley}, {Kolenberg},
  {Kontizas}, {Kontizas}, {Koposov}, {Kordopatis}, {Kostrzewa-Rutkowska},
  {Koubsky}, {Lambert}, {Lanza}, {Lasne}, {Lavigne}, {Le Fustec}, {Le
  Poncin-Lafitte}, {Lebreton}, {Leccia}, {Leclerc}, {Lecoeur-Taibi},
  {Lenhardt}, {Leroux}, {Liao}, {Licata}, {Lindstr{\o}m}, {Lister}, {Livanou},
  {Lobel}, {L{\'o}pez}, {Managau}, {Mann}, {Mantelet}, {Marchal}, {Marchant},
  {Marconi}, {Marinoni}, {Marschalk{\'o}}, {Marshall}, {Martino}, {Marton},
  {Mary}, {Massari}, {Matijevi{\v{c}}}, {Mazeh}, {McMillan}, {Messina},
  {Michalik}, {Millar}, {Molina}, {Molinaro}, {Moln{\'a}r}, {Montegriffo},
  {Mor}, {Morbidelli}, {Morel}, {Morris}, {Mulone}, {Muraveva}, {Musella},
  {Nelemans}, {Nicastro}, {Noval}, {O'Mullane}, {Ord{\'e}novic},
  {Ord{\'o}{\~n}ez-Blanco}, {Osborne}, {Pagani}, {Pagano}, {Pailler},
  {Palacin}, {Palaversa}, {Panahi}, {Pawlak}, {Piersimoni}, {Pineau}, {Plachy},
  {Plum}, {Poggio}, {Poujoulet}, {Pr{\v{s}}a}, {Pulone}, {Racero}, {Ragaini},
  {Rambaux}, {Ramos-Lerate}, {Regibo}, {Reyl{\'e}}, {Riclet}, {Ripepi}, {Riva},
  {Rivard}, {Rixon}, {Roegiers}, {Roelens}, {Romero-G{\'o}mez}, {Rowell},
  {Royer}, {Ruiz-Dern}, {Sadowski}, {Sagrist{\`a} Sell{\'e}s}, {Sahlmann},
  {Salgado}, {Salguero}, {Sanna}, {Santana-Ros}, {Sarasso}, {Savietto},
  {Schultheis}, {Sciacca}, {Segol}, {Segovia}, {S{\'e}gransan}, {Shih},
  {Siltala}, {Silva}, {Smart}, {Smith}, {Solano}, {Solitro}, {Sordo}, {Soria
  Nieto}, {Souchay}, {Spagna}, {Spoto}, {Stampa}, {Steele},
  {Steidelm{\"u}ller}, {Stephenson}, {Stoev}, {Suess}, {Surdej}, {Szabados},
  {Szegedi-Elek}, {Tapiador}, {Taris}, {Tauran}, {Taylor}, {Teixeira},
  {Terrett}, {Teyssandier}, {Thuillot}, {Titarenko}, {Torra Clotet}, {Turon},
  {Ulla}, {Utrilla}, {Uzzi}, {Vaillant}, {Valentini}, {Valette}, {van Elteren},
  {Van Hemelryck}, {van Leeuwen}, {Vaschetto}, {Vecchiato}, {Veljanoski},
  {Viala}, {Vicente}, {Vogt}, {von Essen}, {Voss}, {Votruba}, {Voutsinas},
  {Walmsley}, {Weiler}, {Wertz}, {Wevers}, {Wyrzykowski}, {Yoldas},
  {{\v{Z}}erjal}, {Ziaeepour}, {Zorec}, {Zschocke}, {Zucker}, {Zurbach}, \&
  {Zwitter}}]{GaiaDR2_2018}
{Gaia Collaboration}, {Brown}, A.~G.~A., {Vallenari}, A., {et~al.} 2018, \aap,
  616, A1, \dodoi{10.1051/0004-6361/201833051}

\bibitem[{Harris {et~al.}(2020)Harris, Millman, van~der Walt, Gommers,
  Virtanen, Cournapeau, Wieser, Taylor, Berg, Smith, {et~al.}}]{Harris2020}
Harris, C.~R., Millman, K.~J., van~der Walt, S.~J., {et~al.} 2020, Nature, 585,
  357

\bibitem[{{Helled} {et~al.}(2014){Helled}, {Bodenheimer}, {Podolak}, {Boley},
  {Meru}, {Nayakshin}, {Fortney}, {Mayer}, {Alibert}, \& {Boss}}]{Helled2014}
{Helled}, R., {Bodenheimer}, P., {Podolak}, M., {et~al.} 2014, in Protostars
  and Planets VI, ed. H.~{Beuther}, R.~S. {Klessen}, C.~P. {Dullemond}, \&
  T.~{Henning}, 643, \dodoi{10.2458/azu\_uapress\_9780816531240-ch028}

\bibitem[{{Henning} {et~al.}(2010){Henning}, {Semenov}, {Guilloteau}, {Dutrey},
  {Hersant}, {Wakelam}, {Chapillon}, {Launhardt}, {Pi{\'e}tu}, \&
  {Schreyer}}]{Henning2010}
{Henning}, T., {Semenov}, D., {Guilloteau}, S., {et~al.} 2010, \apj, 714, 1511,
  \dodoi{10.1088/0004-637X/714/2/1511}

\bibitem[{{Huang} \& {{\"O}berg}(2015)}]{Huang2015}
{Huang}, J., \& {{\"O}berg}, K.~I. 2015, \apjl, 809, L26,
  \dodoi{10.1088/2041-8205/809/2/L26}

\bibitem[{Hunter(2007)}]{Hunter2007}
Hunter, J.~D. 2007, Computing in science \& engineering, 9, 90

\bibitem[{{Kama} {et~al.}(2016){Kama}, {Bruderer}, {van Dishoeck},
  {Hogerheijde}, {Folsom}, {Miotello}, {Fedele}, {Belloche}, {G{\"u}sten}, \&
  {Wyrowski}}]{Kama2016}
{Kama}, M., {Bruderer}, S., {van Dishoeck}, E.~F., {et~al.} 2016, \aap, 592,
  A83, \dodoi{10.1051/0004-6361/201526991}

\bibitem[{{Kama} {et~al.}(2020){Kama}, {Trapman}, {Fedele}, {Bruderer},
  {Hogerheijde}, {Miotello}, {van Dishoeck}, {Clarke}, \& {Bergin}}]{Kama2020}
{Kama}, M., {Trapman}, L., {Fedele}, D., {et~al.} 2020, \aap, 634, A88,
  \dodoi{10.1051/0004-6361/201937124}

\bibitem[{{Kenyon} \& {Hartmann}(1987)}]{KenyonHartmann1987}
{Kenyon}, S.~J., \& {Hartmann}, L. 1987, \apj, 323, 714, \dodoi{10.1086/165866}

\bibitem[{{Kim} \& {Turner}(2020)}]{KimTurner2020}
{Kim}, S.~Y., \& {Turner}, N.~J. 2020, \apj, 889, 159,
  \dodoi{10.3847/1538-4357/ab66ae}

\bibitem[{{Krijt} {et~al.}(2020){Krijt}, {Bosman}, {Zhang}, {Schwarz},
  {Ciesla}, \& {Bergin}}]{Krijt2020}
{Krijt}, S., {Bosman}, A.~D., {Zhang}, K., {et~al.} 2020, \apj, 899, 134,
  \dodoi{10.3847/1538-4357/aba75d}

\bibitem[{{Krijt} {et~al.}(2018){Krijt}, {Schwarz}, {Bergin}, \&
  {Ciesla}}]{Krijt2018}
{Krijt}, S., {Schwarz}, K.~R., {Bergin}, E.~A., \& {Ciesla}, F.~J. 2018, \apj,
  864, 78, \dodoi{10.3847/1538-4357/aad69b}

\bibitem[{Long {et~al.}(2017)Long, Herczeg, Pascucci, Drabek-Maunder, Mohanty,
  Testi, Apai, Hendler, Henning, Manara, {et~al.}}]{Long2017}
Long, F., Herczeg, G.~J., Pascucci, I., {et~al.} 2017, \apj, 844, 99

\bibitem[{{McClure} {et~al.}(2016){McClure}, {Bergin}, {Cleeves}, {van
  Dishoeck}, {Blake}, {Evans}, {Green}, {Henning}, {{\"O}berg}, {Pontoppidan},
  \& {Salyk}}]{McClure2016}
{McClure}, M.~K., {Bergin}, E.~A., {Cleeves}, L.~I., {et~al.} 2016, \apj, 831,
  167, \dodoi{10.3847/0004-637X/831/2/167}

\bibitem[{{Miotello} {et~al.}(2014){Miotello}, {Bruderer}, \& {van
  Dishoeck}}]{Miotello2014}
{Miotello}, A., {Bruderer}, S., \& {van Dishoeck}, E.~F. 2014, \aap, 572, A96,
  \dodoi{10.1051/0004-6361/201424712}

\bibitem[{{Miotello} {et~al.}(2016){Miotello}, {van Dishoeck}, {Kama}, \&
  {Bruderer}}]{Miotello2016}
{Miotello}, A., {van Dishoeck}, E.~F., {Kama}, M., \& {Bruderer}, S. 2016,
  \aap, 594, A85, \dodoi{10.1051/0004-6361/201628159}

\bibitem[{{Miotello} {et~al.}(2017){Miotello}, {van Dishoeck}, {Williams},
  {Ansdell}, {Guidi}, {Hogerheijde}, {Manara}, {Tazzari}, {Testi}, {van der
  Marel}, \& {van Terwisga}}]{miotello2017}
{Miotello}, A., {van Dishoeck}, E.~F., {Williams}, J.~P., {et~al.} 2017, \aap,
  599, A113, \dodoi{10.1051/0004-6361/201629556}

\bibitem[{{Mordasini}(2018)}]{Mordasini2018}
{Mordasini}, C. 2018, {Planetary Population Synthesis}, 143,
  \dodoi{10.1007/978-3-319-55333-7_143}

\bibitem[{{Oberg} {et~al.}(2021){Oberg}, {Guzman}, {Walsh}, {Aikawa}, {Bergin},
  {Law}, {Loomis}, {Alarcon}, {Andrews}, {Bae}, {Bergner}, {Boehler}, {Booth},
  {Bosman}, {Calahan}, {Cataldi}, {Cleeves}, {Czekala}, {Furuya}, {Huang},
  {Ilee}, {Kurtovic}, {Le Gal}, {Liu}, {Long}, {Menard}, {Nomura}, {Perez},
  {Qi}, {Schwarz}, {Sierra}, {Teague}, {Tsukagoshi}, {Yamato}, {van 't Hoff},
  {Waggoner}, {Wilner}, \& {Zhang}}]{ObergMAPS2021}
{Oberg}, K.~I., {Guzman}, V.~V., {Walsh}, C., {et~al.} 2021, arXiv e-prints,
  arXiv:2109.06268.
\newblock \doarXiv{2109.06268}

\bibitem[{{Qi} {et~al.}(2015){Qi}, {{\"O}berg}, {Andrews}, {Wilner}, {Bergin},
  {Hughes}, {Hogherheijde}, \& {D'Alessio}}]{Qi2015}
{Qi}, C., {{\"O}berg}, K.~I., {Andrews}, S.~M., {et~al.} 2015, \apj, 813, 128,
  \dodoi{10.1088/0004-637X/813/2/128}

\bibitem[{{Qi} {et~al.}(2013){Qi}, {{\"O}berg}, {Wilner}, {D'Alessio},
  {Bergin}, {Andrews}, {Blake}, {Hogerheijde}, \& {van Dishoeck}}]{Qi2013}
{Qi}, C., {{\"O}berg}, K.~I., {Wilner}, D.~J., {et~al.} 2013, Science, 341,
  630, \dodoi{10.1126/science.1239560}

\bibitem[{{Qi} {et~al.}(2019){Qi}, {{\"O}berg}, {Espaillat}, {Robinson},
  {Andrews}, {Wilner}, {Blake}, {Bergin}, \& {Cleeves}}]{Qi2019}
{Qi}, C., {{\"O}berg}, K.~I., {Espaillat}, C.~C., {et~al.} 2019, \apj, 882,
  160, \dodoi{10.3847/1538-4357/ab35d3}

\bibitem[{{Salinas} {et~al.}(2016){Salinas}, {Hogerheijde}, {Bergin},
  {Cleeves}, {Brinch}, {Blake}, {Lis}, {Melnick}, {Pani{\'c}}, {Pearson},
  {Kristensen}, {Y{\i}ld{\i}z}, \& {van Dishoeck}}]{Salinas2016}
{Salinas}, V.~N., {Hogerheijde}, M.~R., {Bergin}, E.~A., {et~al.} 2016, \aap,
  591, A122, \dodoi{10.1051/0004-6361/201628172}

\bibitem[{{Schwarz} {et~al.}(2016){Schwarz}, {Bergin}, {Cleeves}, {Blake},
  {Zhang}, {{\"O}berg}, {van Dishoeck}, \& {Qi}}]{Schwarz2016}
{Schwarz}, K.~R., {Bergin}, E.~A., {Cleeves}, L.~I., {et~al.} 2016, \apj, 823,
  91, \dodoi{10.3847/0004-637X/823/2/91}

\bibitem[{{Schwarz} {et~al.}(2018){Schwarz}, {Bergin}, {Cleeves}, {Zhang},
  {{\"O}berg}, {Blake}, \& {Anderson}}]{Schwarz2018}
---. 2018, \apj, 856, 85, \dodoi{10.3847/1538-4357/aaae08}

\bibitem[{{Schwarz} {et~al.}(2019){Schwarz}, {Bergin}, {Cleeves}, {Zhang},
  {{\"O}berg}, {Blake}, \& {Anderson}}]{Schwarz2019}
---. 2019, \apj, 877, 131, \dodoi{10.3847/1538-4357/ab1c5e}

\bibitem[{{Schwarz} {et~al.}(2021){Schwarz}, {Calahan}, {Zhang}, {Alarc{\'o}n},
  {Aikawa}, {Andrews}, {Bae}, {Bergin}, {Booth}, {Bosman}, {Cataldi},
  {Cleeves}, {Czekala}, {Huang}, {Ilee}, {Law}, {Le Gal}, {Liu}, {Long},
  {Loomis}, {Mac{\'\i}as}, {McClure}, {M{\'e}nard}, {{\"O}berg}, {Teague}, {van
  Dishoeck}, {Walsh}, \& {Wilner}}]{SchwarzMAPS2021}
{Schwarz}, K.~R., {Calahan}, J.~K., {Zhang}, K., {et~al.} 2021, \apjs, 257, 20,
  \dodoi{10.3847/1538-4365/ac143b}

\bibitem[{{Seifert} {et~al.}(2021){Seifert}, {Cleeves}, {Adams}, \&
  {Li}}]{Seifert2021}
{Seifert}, R.~A., {Cleeves}, L.~I., {Adams}, F.~C., \& {Li}, Z.-Y. 2021, \apj,
  912, 136, \dodoi{10.3847/1538-4357/abf09a}

\bibitem[{{Thi} {et~al.}(2010){Thi}, {Mathews}, {M{\'e}nard}, {Woitke},
  {Meeus}, {Riviere-Marichalar}, {Pinte}, {Howard}, {Roberge}, {Sandell},
  {Pascucci}, {Riaz}, {Grady}, {Dent}, {Kamp}, {Duch{\^e}ne}, {Augereau},
  {Pantin}, {Vandenbussche}, {Tilling}, {Williams}, {Eiroa}, {Barrado},
  {Alacid}, {Andrews}, {Ardila}, {Aresu}, {Brittain}, {Ciardi}, {Danchi},
  {Fedele}, {de Gregorio-Monsalvo}, {Heras}, {Huelamo}, {Krivov}, {Lebreton},
  {Liseau}, {Martin-Zaidi}, {Mendigut{\'{\i}}a}, {Montesinos}, {Mora},
  {Morales-Calderon}, {Nomura}, {Phillips}, {Podio}, {Poelman}, {Ramsay},
  {Rice}, {Solano}, {Walker}, {White}, \& {Wright}}]{Thi2010}
{Thi}, W.-F., {Mathews}, G., {M{\'e}nard}, F., {et~al.} 2010, \aap, 518, L125,
  \dodoi{10.1051/0004-6361/201014578}

\bibitem[{{Toomre}(1964)}]{Toomre1964}
{Toomre}, A. 1964, \apj, 139, 1217, \dodoi{10.1086/147861}

\bibitem[{{Trapman} {et~al.}(2017){Trapman}, {Miotello}, {Kama}, {van
  Dishoeck}, \& {Bruderer}}]{Trapman2017}
{Trapman}, L., {Miotello}, A., {Kama}, M., {van Dishoeck}, E.~F., \&
  {Bruderer}, S. 2017, \aap, 605, A69, \dodoi{10.1051/0004-6361/201630308}

\bibitem[{{van Dishoeck} \& {Black}(1988)}]{vanDishoeckBlack1988}
{van Dishoeck}, E.~F., \& {Black}, J.~H. 1988, \apj, 334, 771,
  \dodoi{10.1086/166877}

\bibitem[{{van 't Hoff} {et~al.}(2017){van 't Hoff}, {Walsh}, {Kama},
  {Facchini}, \& {van Dishoeck}}]{vtHoff2017}
{van 't Hoff}, M.~L.~R., {Walsh}, C., {Kama}, M., {Facchini}, S., \& {van
  Dishoeck}, E.~F. 2017, \aap, 599, A101, \dodoi{10.1051/0004-6361/201629452}

\bibitem[{{van Zadelhoff} {et~al.}(2001){van Zadelhoff}, {van Dishoeck}, {Thi},
  \& {Blake}}]{Zadelhoff2001}
{van Zadelhoff}, G.~J., {van Dishoeck}, E.~F., {Thi}, W.~F., \& {Blake}, G.~A.
  2001, \aap, 377, 566, \dodoi{10.1051/0004-6361:20011137}

\bibitem[{{Visser} {et~al.}(2009){Visser}, {van Dishoeck}, \&
  {Black}}]{Visser2009}
{Visser}, R., {van Dishoeck}, E.~F., \& {Black}, J.~H. 2009, \aap, 503, 323,
  \dodoi{10.1051/0004-6361/200912129}

\bibitem[{{Williams} \& {Best}(2014)}]{WilliamsBest2014}
{Williams}, J.~P., \& {Best}, W.~M.~J. 2014, \apj, 788, 59,
  \dodoi{10.1088/0004-637X/788/1/59}

\bibitem[{{Yu} {et~al.}(2017){Yu}, {Evans}, {Dodson-Robinson}, {Willacy}, \&
  {Turner}}]{Yu2017}
{Yu}, M., {Evans}, Neal~J., I., {Dodson-Robinson}, S.~E., {Willacy}, K., \&
  {Turner}, N.~J. 2017, \apj, 841, 39, \dodoi{10.3847/1538-4357/aa6e4c}

\bibitem[{{Yu} {et~al.}(2016){Yu}, {Willacy}, {Dodson-Robinson}, {Turner}, \&
  {Evans}}]{Yu2016}
{Yu}, M., {Willacy}, K., {Dodson-Robinson}, S.~E., {Turner}, N.~J., \& {Evans},
  Neal~J., I. 2016, \apj, 822, 53, \dodoi{10.3847/0004-637X/822/1/53}

\bibitem[{{Zhang} {et~al.}(2017){Zhang}, {Bergin}, {Blake}, {Cleeves}, \&
  {Schwarz}}]{Zhang2017}
{Zhang}, K., {Bergin}, E.~A., {Blake}, G.~A., {Cleeves}, L.~I., \& {Schwarz},
  K.~R. 2017, Nature Astronomy, 1, 0130, \dodoi{10.1038/s41550-017-0130}

\bibitem[{{Zhang} {et~al.}(2019){Zhang}, {Bergin}, {Schwarz}, {Krijt}, \&
  {Ciesla}}]{Zhang2019}
{Zhang}, K., {Bergin}, E.~A., {Schwarz}, K., {Krijt}, S., \& {Ciesla}, F. 2019,
  \apj, 883, 98, \dodoi{10.3847/1538-4357/ab38b9}

\bibitem[{{Zhang} {et~al.}(2021){Zhang}, {Booth}, {Law}, {Bosman}, {Schwarz},
  {Bergin}, {{\"O}berg}, {Andrews}, {Guzm{\'a}n}, {Walsh}, {Qi}, {van't Hoff},
  {Long}, {Wilner}, {Huang}, {Czekala}, {Ilee}, {Cataldi}, {Bergner}, {Aikawa},
  {Teague}, {Bae}, {Loomis}, {Calahan}, {Alarc{\'o}n}, {M{\'e}nard}, {Le Gal},
  {Sierra}, {Yamato}, {Nomura}, {Tsukagoshi}, {P{\'e}rez}, {Trapman}, {Liu}, \&
  {Furuya}}]{ZhangMAPS2021}
{Zhang}, K., {Booth}, A.~S., {Law}, C.~J., {et~al.} 2021, \apjs, 257, 5,
  \dodoi{10.3847/1538-4365/ac1580}

\end{thebibliography}

\begin{appendix}

\section{Gas Mass fits based on HD $J=1-0$}
\label{app: HD gas masses}

\begin{figure}
    \centering
    \includegraphics[width=\columnwidth]{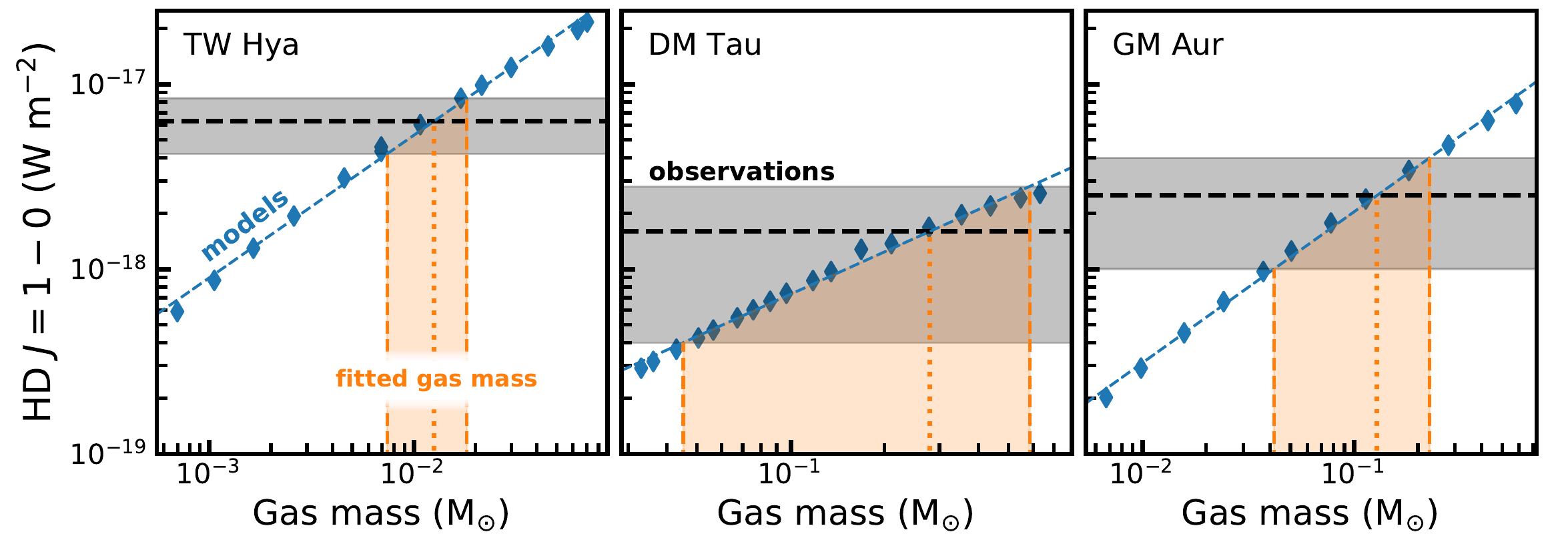}
    \caption{\label{fig: HD gas masses} Gas mass ranges of TW Hya, DM Tau and GM Aur obtained by comparing model HD 1-0 integrated fluxes to the observations \citep{Bergin2013, McClure2016}. Observations are shown in black, with the gray shaded region showing the $3\sigma$ uncertainty. Blue points show individual model fluxes, to which we have fitted the power law shown as the blue dashed line. The orange vertical line and shaded region show the best fitting gas mass and $3\sigma$ uncertainty, respectively. }
\end{figure}

In Figure \ref{fig: HD gas masses} we show how gas masses were measured from the HD 1-0 integrated line flux.

\end{appendix}

\end{document}